\documentclass[prl,aps,twocolumn,twoside,english,superscriptaddress]{revtex4}

\usepackage{epsfig}
\usepackage{graphicx}
\usepackage{amsmath}
\usepackage{revsymb}
\newcommand{\ket}[1]{|  #1 \rangle}

\newcommand\bea{\begin{eqnarray}}
\newcommand\eea{\end{eqnarray}}
\newcommand{\beq}{\begin{equation}}
\newcommand{\eeq}{\end{equation}}

\begin{document}

\title{\Large Inequalities and Separations among Assisted Capacities
of Quantum Channels}
\author{Charles H. Bennett}\email{bennetc@watson.ibm.com}
\affiliation{IBM T.J. Watson Research Center, Yorktown Heights, NY
10598, USA}
\author{Igor Devetak}\email{devetak@csi.usc.edu}
\affiliation{IBM T.J. Watson Research Center, Yorktown Heights, NY
10598, USA}
\author{Peter W. Shor}\email{shor@math.mit.edu}
\affiliation{Massachusetts Institute of Technology, Cambridge, MA
02139, USA}
\author{John A. Smolin}\email{smolin@watson.ibm.com}
\affiliation{IBM T.J. Watson
Research Center, Yorktown Heights, NY 10598, USA}

\date{20 February 2004}

\begin{abstract}
We exhibit discrete memoryless quantum channels whose quantum
capacity assisted by two-way classical communication, $Q_2$,
exceeds their unassisted one-shot Holevo capacity $C_H$. These
channels may be thought of as having a data input and output,
along with a control input that partly influences, and a control
output that partly reveals, which of a set of unitary evolutions
the data undergoes en route from input to output. The channel is
designed so that the data's evolution can be exactly inferred by a
classically coordinated processing of 1) the control output, and
2) a reference system entangled with the control input, but not
from either of these resources alone. Thus a two-way classical
side channel allows the otherwise noisy evolution of the data to
be corrected, greatly increasing the capacity. The same family of
channels provides examples where the classical capacity assisted
by classical feedback, $C_{B}$, and the quantum capacity assisted
by classical feedback $Q_{B}$, both exceed $C_H$.  A related
channel, whose data input undergoes dephasing before interacting
with the control input, has a classical capacity $C=C_H$ strictly
less than its $C_2$, the classical capacity assisted by
independent classical communication.
\end{abstract}

\maketitle

\subsection{Introduction} \label{sec:intro}

Perhaps the most important open question concerning quantum
discrete memoryless channels (DMCs) is whether, as is widely
believed, the asymptotic classical capacity $C$ is equal to the
one-shot Holevo capacity $C_H$, defined as the maximum, over input
distributions for a single use of the channel, of the entropy of
the average output minus the average of the output
entropies~\cite{HSW,Add}. Another open question is whether
classical feedback can increase a quantum channel's classical
capacity. For classical DMCs it has long been known that feedback
does not increase the capacity, and Bowen et al~\cite{BN} recently
showed that this is also true for entanglement-breaking quantum
DMCs, and for the entanglement-assisted classical capacity of any
quantum DMC~\cite{Bowen}. A related open question is whether there
is any channel whose quantum capacity assisted by two-way
classical communication $Q_2$, or by classical feedback, $Q_{B}$,
exceeds the unassisted classical capacity $C$.  Here we exhibit
channels for which $C_{B}$, $Q_2$ and $Q_{B}$ all exceed $C_H$
(and therefore $C$, if $C_H\!=\!C$).

To achieve these separations we use a special kind of quantum
discrete memoryless channel which we call {\em retrocorrectable}.
It may be thought of as having a data input and output, along with
a control input that partly influences, and a control output that
partly reveals, which of a set of unitary evolutions the data has
undergone en route from input to output. The channel is designed
so that the data's evolution can be exactly inferred by a
classically coordinated processing of 1) the control output, and
2) a reference system entangled with the control input, but not
from either of these resources alone. Thus a two-way classical
side channel allows the data's otherwise noisy evolution to be
corrected, greatly increasing the capacity. These channels are
described in section 1 and applied in sections 2 and 3.

The channels used in these constructions are not
entanglement-breaking, and are not known to have $C=C_H$.  However
in section 4 we consider a weaker kind of retrocorrectable
channel, in which the data input undergoes complete dephasing
before interacting with the control input.  These channels are
entanglement-breaking by construction and therefore have
$C_{B}=C=C_H$ and $Q_2=0$. Nevertheless, they have the
nonclassical feature that their classical capacity is strictly
less than their $C_2$, the classical capacity assisted by two-way
classical communication independent of the message to be
transmitted.  Section 5 discusses other results and remaining open
questions concerning relations among the capacities.

\subsection{1. The echo effect and retrocorrectable
channels}

Before introducing the class of channels that we will be using to
prove capacity separations, we consider the underlying phenomenon
in a simpler setting, that of a classic Bell-inequality
experiment.  In a typical such experiment, each member of a pair
of polarization-entangled photons enters a separate analyzer which
chooses randomly, and independently of the other analyzer, one of
two nonorthogonal bases in which to measure the photon's
polarization. The output of the analyzer is two classical bits
indicating the basis $b$, and the result $j$ of measuring the
photon in that basis. The analyzer may be viewed abstractly as a
quantum-classical (QC) channel taking a single quantum input to
two classical outputs. Given such a two-output QC channel, we say
that one of its outputs, here $j$, is externally {\em echoed\/} if
it can be accurately inferred from the other output (here $b$) and
a system entangled with the channel input (in this case the other
photon of the entangled pair), but could not have been accurately
inferred had the input instead been a known unentangled pure
state.  The fact that supplying an input photon of known
polarization is worse than supplying one with an entangled partner
to be measured later is an essential manifestation of entanglement
and the violations of Bell's inequality it gives rise to.

In our applications the externally echoed quantity, instead of
being emitted as an output, is used internally within the channel
to control the processing of another input.  In more detail, we
consider channels whose input and output spaces are each
conceptually factorized into a control part and a data part.  The
channel performs a stochastic mapping of the data input variable
onto a corresponding output variable, in a way that is partly
influenced by the control input, and is partly revealed by the
control output. The goal is to design the channel so that the
stochastic mapping of the data variable can be corrected and made
noiseless with the help of measurements on the control output and
on a reference system entangled with the control input, but not by
either resource alone. Two-way communication allows the
measurements at the sending and receiving ends to be coordinated
and exploited, thereby increasing the capacity above what could be
achieved by any noninteractive protocol.  Of course it is
necessary to be sure that the control input and output do not, by
their mere presence, increase the Holevo capacity so much as to
neutralize the gains achieved by using them to correct the data
variable's stochastic evolution.

Specifically we consider a family of channels, for integers
$c,d\geq 2$, which we call {\em standard retrocorrectable
channels\/} and denote ${\cal R}_{c,d}$.  Their internal operation
is depicted in Figure 1.

\begin{figure}
\includegraphics[width=8cm]{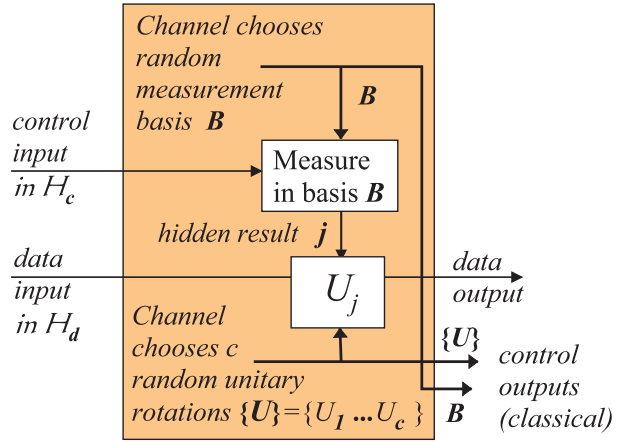}
\caption{The standard retrocorrectable channel ${\cal R}_{c,d}$
has a data input living in a $d$-dimensional Hilbert space, and a
corresponding output; a control input living in a $c$-dimensional
Hilbert space; and a classical control output living in a large
Hilbert space. The channel internally selects a random basis $B$,
for ${\cal H}_{c}$ and a set of $c$ random unitaries
$\{U\}=U_1...U_c$ on ${\cal H}_d$. The channel measures the
control input in the basis $B$, yielding result $j\in\{1...c\}$.
and according to that result applies one of the unitaries $U_j$ to
the data input, which is then emitted as the data output. The
channel also emits a classical control output consisting of the
random basis $B$ and the set of random unitaries
$\{U\}=U_1...U_c$.  It does not, however, emit the measurement
result $j$ but keeps it hidden, or, in another way of speaking,
discards it into the channel's inaccessible environment.}
\label{chandef}
\end{figure}

\subsection{2. Holevo capacity and coherent information} \label{sec:Hcap}
Owing to the retrocorrectable channels' high symmetry, their
one-shot Holevo capacity can be calculated assuming a uniform
distribution over the data input and an arbitrary fixed value of
the control input. The control outputs do not contribute to the
Holevo capacity because they are uncorrelated with the input. In
the $c\!=\!d\!=\!2$ (qubit) case \beq C_H=1+(\pi^2\!/18-5/6)/\ln2
= 0.5888 \eeq.

If we allow $c$ to increase slightly superlinearly with $d$ (e.g.
as $c=d\log^3d$), the randomization of the data becomes more
efficient for larger $d$, making $C_H$ tend to zero as
$d\!\rightarrow\!\infty$. This follows the fact that under these
conditions, 1) all but an asymptotically vanishing fraction of the
probability distribution of outcomes $j$ is contributed by
outcomes of probability less than $1/(d\log^2d)$~\cite{W90}, and
2) except for these improbable outcomes, the application of an
unknown unitary from a known set of random unitaries on ${\cal
H}_d$ of cardinality $d\log^2 d$ constitutes an asymptotically
randomizing quantum channel in the sense of Hayden et.
al.~\cite{HLSW0307104}.

The one-shot coherent information is similarly maximized by
choosing a uniform ensemble for the data variable and a fixed
value for the control variable.  Although the one-shot coherent
information may be only a lower bound on the unassisted quantum
capacity for this channel, the latter is in any case bounded above
by the unassisted classical capacity $C$, and so, if $C_H$ is
additive, by $C_H$.   The maximal one-shot coherent information is
0.4262 for $c\!=\!d\!=\!2$ and approaches zero for
$d\!\rightarrow\!\infty$, if, as before, $c$ is allowed to
increase slightly superlinearly with $d$.

\subsection{3. Assisted Capacities} \label{asscap}

By using the echo effect to retrospectively correct the otherwise
noisy evolution of the data qubit, we can show that for
appropriate choices of $c$ and $d$, the retrocorrectable channel's
channel's assisted capacities, $C_{B}$, $Q_2$, $C_2$, $Q_E$ and
$C_E$ all can be made to exceed the one-shot Holevo capacity
$C_H$.  Indeed, by allowing $c$ to increase slightly superlinearly
with $d$ as in the previous section, all the assisted capacities
can be made to increase linearly with $\log d$, while $C_H$ tends
to zero.

Figures 2-5 show respectively how the channel ${\cal R}_{2,2}$ can
be used \begin{itemize}
\item to transmit a faithful qubit in the presence of two-way
communication;
\item to generate a faithful ebit in the presence of classical back communication;
and
\item to transmit a faithful qubit without back communication, but
consuming an ebit previously shared between sender and receiver.
\end{itemize}

The essential trick, shown in Figure 2, is to feed the control
input half of an ebit (a maximally entangled pair of qubits), and
then to measure the other half in the basis $B^T$, thereby
creating an echo of the measurement result $j$ that had been
obtained earlier but discarded within the channel. In figure 2,
Alice performs this measurement after she has learned the value of
$B$ through the back channel.  She then sends the measurement
result to Bob, which allows him to correct the evolution of the
data qubit, resulting in a faithful qubit.
\begin{figure}
\includegraphics[width=8cm]{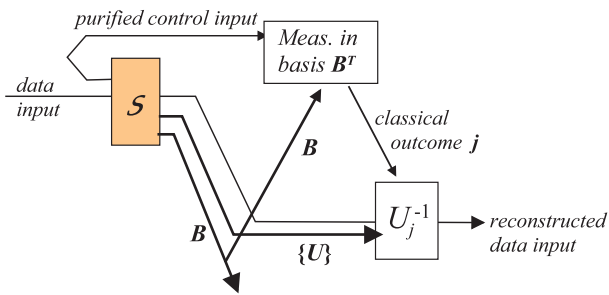}
\caption{Protocol for using the retrocorrectable channel ${\cal
S}={\cal R}_{2,2}$, in conjunction with two-way classical
communication, to implement a noiseless qubit channel. Sender
Alice is at the top, receiver Bob at the bottom. Alice feeds the
control input half of a maximally entangled pair, whose other half
she later measures in the basis $B^T$, after Bob has told her $B$
through a classical back channel.  This measurement, yields, via
the ``echo effect," the same outcome $j$ as occurred earlier
within the channel. Alice tells Bob $j$ through a forward
classical channel, after which he can undo the unitary
transformation $U_j$ that the channel performed, restoring the
data output qubit to the same state as it had initially.}
\end{figure}

Figure 3 shows how the channel ${\cal S}$ together with back
communication can be used to create an ebit. Finally, figure 4
shows how the channel, in conjunction with an ebit shared earlier,
can be used to implement a faithful forward qubit channel.

\begin{figure}
\includegraphics[width=8cm]{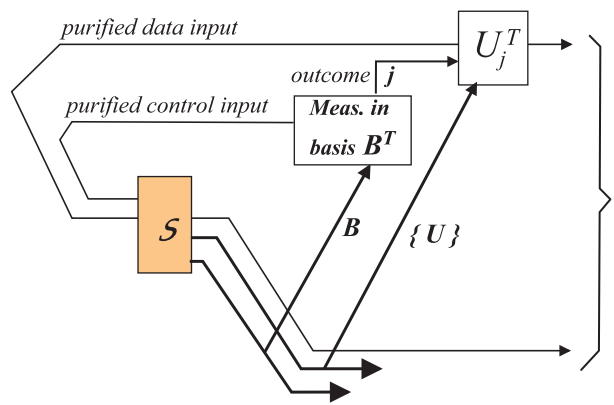}
\caption{When used in conjunction with backward classical
communication, the (2,2) retrocorrectable channel ${\cal S}$ can
be used to generate an a standard maximally entangled pair of
qubits. To do this, Alice feeds both inputs with halves of
maximally entangled states. The auxiliary input's reference system
is used in conjunction with a classical back channel as before to
learn the internal measurement outcome $j$ through the echo
effect. Alice then applies the unitary transformation $U_j^T$ to
the reference system entangled with the main input, thereby
restoring the combination of it and the main output to a standard
maximally entangled state.}
\end{figure}

\begin{figure}
\includegraphics[width=8cm]{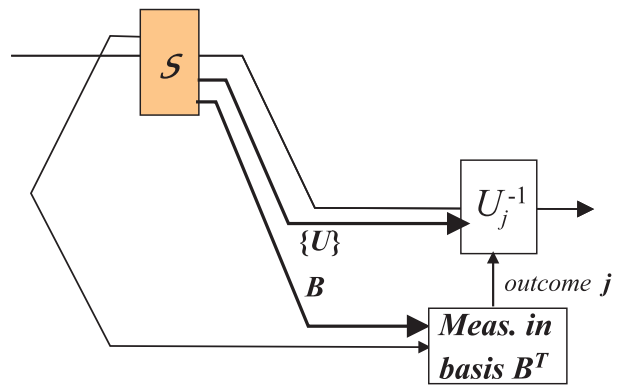}
\caption{When used in conjunction with an ebit shared beforehand
between Alice and Bob, the channel ${\cal S}$ can be used to
generate a faithful qubit channel.}
\end{figure}

Continuing to use ${\cal S}$ to represent the standard
retrocorrectable channel ${\cal R}_{2,2}$ of Figure 1, we have the
following reducibilities.
\begin{eqnarray}
1\;\; {\rm ebit}\;\; & \leq  & \;\;{\cal S} + \mbox{\rm back
communication}\\
1\;\; {\rm qubit}\;\; & \leq & \;\;{\cal S}\; +1\;\; {\rm ebit}
\end{eqnarray}
These imply
\[
1\;\; {\rm qubit}\;\; \leq \;\;2\;{\cal S}\;+ \mbox{\rm back
communication}
\]
and, using superdense coding,
\[
2\;\; {\rm cbit}\;\; \leq \;\;3\; {\cal S}\; + \mbox{\rm back
communication}
\]

These constructions may be extended to variable $c$ and $d$, and,
in the limit of large $d$ with $c$ increasing slightly
superlinearly with $d$, we have the following:
\begin{eqnarray}
\lim_{d\rightarrow\infty} C_H &=& 0 \\
\lim_{d\rightarrow\infty}Q_2 &=&  \log d \\
\lim_{d\rightarrow\infty}Q_{B} &\geq& \frac{1}{2}\log d \\
\lim_{d\rightarrow\infty}C_{B} &\geq& \frac{2}{3} \log d
\end{eqnarray}

The last two expressions are lower bounds, because it is possible
(though we have no evidence for it) that a higher $Q_B$ or $C_B$
might be achieved by some other protocol than ours.

\subsection{4. Other similar channels}\label{similarchans}

Other simpler channels with nontrivial echo effects can be
constructed.  For example, consider a simplified (2,2) channel
that measures its control qubit in a random one of two fixed
conjugate bases and, according to result, either does or doesn't
depolarize its data qubit input before emitting it as the data
output. The control output is then a single classical bit
indicating the measurement basis. While the limited control
information does not allow the data variable's evolution to be
corrected completely, it does allow depolarization events to be
converted to less costly erasures, thereby creating a separation
between $Q_2$ and $C_H$. The maximum $Q_2$ of
$\cos^2\frac{\pi}{8}=0.85355$ is achieved by setting the control
qubit midway between the $\ket{0}$ eigenstates of the two
conjugate bases, while the maximum $C_H$ of
$\frac{1}{2}+\frac{1}{2}(1-h_2(\frac{1}{4}))=0.59436$ is obtained
by setting it equal to the $\ket{0}$ eigenstate of one of them.
This channel is the simplest example of what might be called a
partially retrocorrectable channel.

The channels considered so far are not entanglement-breaking. To
obtain an echo effect in an entanglement-breaking channel, we
modify the the standard (2,2) retrocorrectable channel ${\cal
S}={\cal R}_{2,2}$ by unconditionally dephasing its data input in
the computational basis before applying the conditional unitary.
Being entanglement-breaking, this channel can have no quantum
capacity without entanglement assistance $(Q=Q_B=Q_2=0)$, but it
does have a $C_2$ capacity of 1, strictly greater than its Holevo
capacity of 0.5888. Since it is entanglement-breaking, its $C_B$
capacity must also equal the Holevo capacity, by the argument of
Bowen and Nagarajan~\cite{BN}. Therefore this channel definitely
(without any assumptions about additivity of $C_H$) violates the
second part of the equality $C=C_B=C_2$ obeyed by all classical
DMCs.

\subsection{5. Relations among capacities}
Assuming for the remainder of this section the additivity of
Holevo capacity, we can construct a ladder diagram (Fig.~5)
showing a double hierarchy with classical capacities $C\leq
C_B\leq C_2 \leq C_E$ on one side and corresponding quantum
capacities $Q\leq Q_B\leq Q_2 \leq Q_E$ on the other.  Each
quantum capacity is upper bounded by its corresponding classical
capacity, and in every case but $Q_E$ vs $C_E$, the inequality can
be saturated. Here $C_2$ denotes a channel's classical capacity
when assisted by an arbitrary classical two-way side conversation,
subject only to the limitation that, taken as a whole, the side
conversation be independent of the message being transmitted
through the main protocol.  With this restriction it is easy to
show that $C_2=C$ for any classical DMC, whereas without it the
side conversation would become a short circuit making $C_2$
trivially infinite.  It has not been customary to impose a similar
independence restriction in the definition of $Q_2$, where no
short circuit problem exists.  But in fact, without loss of
generality, the classical side conversation in $Q_2$ can also be
required to be independent of the (quantum) message being
transmitted through the main protocol, because if it were not
independent, it could be used as a means of eavesdropping on the
quantum message without disturbing it. From another viewpoint,
$C_2$ represents a channel's private classical capacity when
assisted by two-way public communication, the adversary being
given access to the side conversation but not the channel
environment (cf~\cite{D0304127}).

In passing, we note that attempting to define capacities such as
``$Q_{E2}$'' which would allow unlimited amounts of {\em both\/}
shared entanglement {\em and\/} bidirectional classical
communication, leads to a more serious short circuit problem,
because independence does not prevent the assistive resources from
being used for teleportation, making the capacity infinite.

Returning to Fig.~5, the general goal is to determine, for every
pair of capacities, whether they are related by
\begin{itemize}
\item a strict inequality, as in the case of $Q_E<C_E$, with the
inequality being saturated only trivially when both sides vanish;
\item a saturable inequality as in $Q\leq C$; or
\item an incomparability as between $Q_E$ and $C$, in which,
depending on the channel, either side may be greater.
\end{itemize}
The former two relations are indicated by a solid line in the
ladder diagram, with the greater quantity being higher.
Incomparability is indicated by a dashed line.

\begin{figure}
\includegraphics[width=8cm]{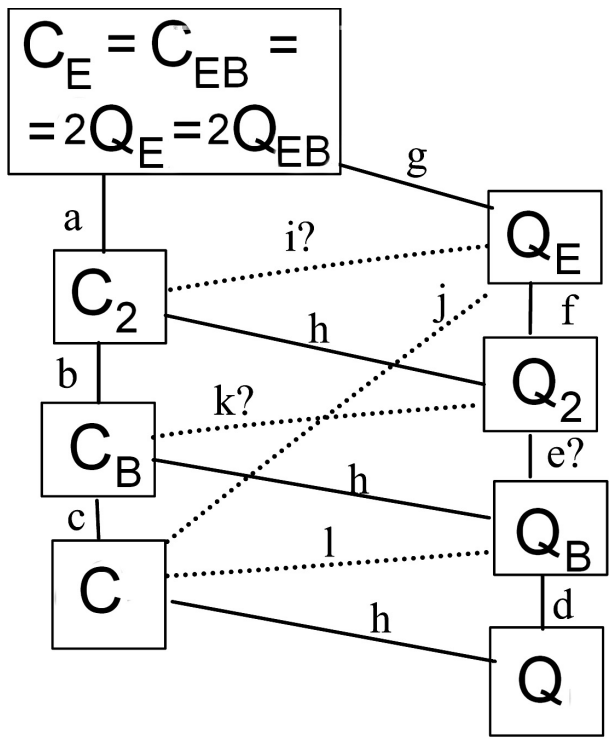}
\caption{Inequalities among capacities. A solid line denotes an
inequality with the higher side being strictly greater than the
lower side for some channels, but equal for other channels. A
dotted line denotes an incomparability, where depending on the
channel either capacity may be greater. The equalities in top left
box follow from Bowen\cite{Bowen}, along with teleportation and
superdense coding. See text for comments labelled {\bf a...l}.
Question marks indicate instances where one or more aspects of the
relation are conjectural.}
\end{figure}

For each of the solid lines in Fig.~5, we need to find a proof of
the general inequality and examples showing both equality and
separation. Referring to the notes {\bf a} through {\bf h} in the
diagram,

\medskip
\noindent{\bf a:} The general inequality  $\leq$ can be shown by
an argument involving monotonicity of the conditional mutual
information (cf \cite{CW}). Equality is witnessed by the
classical bit channel (ie a 100\% dephasing qubit channel), for
which it is easy to show that both capacities equal 1. Separation
is witnessed by the noiseless qubit channel where we can show that
$C_2=1$ but by superdense coding $C_E=2$.

\noindent{\bf b:} The general inequality follows from the fact
that any protocol that achieves $C_B$ can be modified to decouple
the back communication from the message (cf \cite{SW}).  Equality
is witnessed by the classical bit channel, inequality by the
dephased retrocorrectable channel of the previous section, for
which $C_2\!=\!1$ but $C_B\!=\!C\!=\!C_H\!<\!1$.

\noindent{\bf c:} The general inequality is obvious.  Equality
holds for the classical bit channel, separation for the standard
retrocorrectable channel (assuming $C=C_H$).

\noindent{\bf d:}  The general inequality is obvious.  Equality
holds for the qubit channel, separation for appropriate high
dimensional retrocorrectable channels.

\noindent{\bf e:}   The general inequality is obvious.  Equality
holds for the qubit channel.  We suspect but do not know how to
prove that $Q_B<Q_2$ for channels such as ${\cal R}_{2,2}$. This
separation would be implied by incomparability {\bf k} which we
also don't know how to prove.

\noindent{\bf f:}  The general inequality can be proved by a
monotonicity argument similar to note {\bf a} above.  Equality is
witnessed by the qubit channel, separation by channels such as the
strongly depolarizing channel, for which $Q_2$ is zero but $C$ and
hence $Q_E$ are positive.

\noindent{\bf g:} These capacities are related by a constant
factor of 2, making inequality strict unless both capacities
vanish.

\noindent{\bf h:} The general inequalities are obvious, from the
fact that a qubit channel can simulate a bit channel.  Separations
are witnessed by the classical bit channel.
\medskip

Turning now to the incomparabilities in Fig. 5,

\medskip
\noindent{\bf i:} $C_2<Q_E$ may be witnessed by the 2/3
depolarizing channel. This channel is known to have $Q_E>C_H$, and
because it is unital, $C=C_H$.  If we can show that $C_2=C_H$ for
this channel then we have a separation. The other separation
$C_2>Q_E$ is witnessed by the classical bit channel.

\noindent{\bf j:} $C<Q_E$ is witnessed by ${\cal R}_{2,2}$
retrocorrectable channel (assuming additivity of $C_H$).  $C>Q_E$
is witnessed by classical bit channel.

\noindent{\bf k:} $C_B>Q_2$ is witnessed by classical channel for
which $Q_2=0$. We conjecture that the retrocorrectable channel
${\cal R}_{2,2}$, with $Q_2=1$ and $C_B\geq 1/2$ witnesses the
inequality in the other direction, but we do not have a nontrivial
upper bound on $C_B$ for this channel.

\noindent{\bf l:} $C<Q_B$ holds for high dimensional
retrocorrectable channels, assuming additivity of $C_H$.  In the
other direction, $C>Q_B$ is witnessed by the classical bit
channel, for which $Q_2=Q_B=0$.

\subsection*{Acknowledgments}

We thank Bill Wootters for helpful discussions of random quantum
states. CHB, ID, and JAS thank the US National Security Agency and
the Advanced Research and Development Activity for support through
contract DAAD19-01-C-0056.

\end{document}